# Neutron Reflection from the Surface of Liquid $^4$He with and without a Layer of $^3$He

T. R. Charlton[1], R. M. Dalgliesh[1], A. Ganshin[2], O. Kirichek[1], S. Langridge[1], P. V. E. McClintock[2]

[1] *ISIS, STFC, Rutherford Appleton Laboratory, Harwell, Didcot, UK*
[2] *Physics Department, Lancaster University, Lancaster, LA1 4YB, UK*

**We report and discuss the first neutron reflection measurements from the free surface of normal and superfluid $^4$He and of liquid $^3$He-$^4$He mixture. In case of liquid $^4$He the surface roughness is different above and below the lambda transition, being smoother in the superfluid state. For the superfluid, we also observe the formation of a surface layer ~200 Å thick which has a subtly different neutron scattering cross-section. The results can be interpreted as an enhancement of Bose-Einstein condensate fraction close to the helium surface. We find that the addition of $^3$He isotopic impurities leads to the formation of Andreev levels at low temperatures.**

## 1. Introduction

The surface properties of the quantum fluids have been the subject of intensive investigation over many years [1 – 4, 22]. Interest in these systems lies both in the fundamental properties of quantum surfaces and also in the opportunity to use them as test media. For example fundamental aspects of the theory of wave turbulence have been tested experimentally [5]. This latter aspect is important because it is very difficult to obtain good experiential results from observations of natural turbulent events, e.g. in the atmosphere, channels, or streams. Furthermore, it is typically very difficult or impossible to control the experimental parameters effectively. Hence there is very strong need for a model system where fundamental aspects of the theory can be tested through laboratory-scale experiments. Another important issue to be investigated is the surface ripplons, i.e. quantized gravity-capillary waves on the free surface of the liquid, or at the interface between two superfluids [6, 21]. For example there is a chance to obtain unique information about the superfluid Kelvin-Helmholtz instability and constitute a model for measurements in the study of the instability of the quantum vacuum beyond the event horizon and in the ergoregion (where the ripplon energy will be negative) of black holes [7]. Up to now, measurements on two-dimensional charge systems including surface electrons (SE) and layer of ions trapped under the surface of liquid helium [8] have provided what is probably the most powerful experimental technique for studying the surface properties of liquid helium. The existence of the quantized capillary surface waves, or ripplons, originally proposed by Atkins [9] in order to explain surface tension data, was clearly proven in experiments on two-dimensional plasma resonance in SE [10], and in the complete control regime [11]. Later, ripplons were detected on liquid $^3$He [12] as well as scattering of SE with Fermi quasi-particles from bulk liquid [13]. Soon after that, it was found that the mobility of SE on $^3$He deviates drastically from single-electron-ripplon scattering theory below 70 mK [14]. However, the SE system provides a rather indirect way of studying the liquid helium surface experimentally, inevitably leading to ambiguities in the interpretation of the data.

In this paper we present the initial results of a new experimental method for studying ripplons and other surface properties, based on neutron scattering. It exploits the unique combination of small-angle neutron reflection from the liquid surface and ultra-low temperature sample environment [15], thus opening new opportunities for studying the interface properties of quantum liquids. We present our measurements of reflection from the superfluid and normal liquid $^4$He surfaces, and preliminary results from dilute $^3$He-$^4$He mixtures.

## 2. Experimental details

Our experiments were performed on the CRISP Instrument at ISIS/RAL, which was designed as a general purpose reflectometer for the investigation of a wide spectrum of interfaces and surfaces. It uses a broad-band neutron time-of-flight method for determination of the wavelength at fixed angles. Typical neutron wavelengths is in the range between 0.5 - 6.5 Å at the source frequency of 50 Hz. To provide the cryogenic sample environment we used a Variox$^{BL}$ cryostat, whose lowest temperature of 1.25K was farther reduced by use of a Heliox insert to cover the temperature range 0.3K to 1.5K.

## 3. Results obtained from pure $^4$He

The measured reflectivity of the liquid $^4$He surface as a function of $Q_z$ at temperatures of 2.3 K and 1.54 K is presented in Fig. 1. Based on the Born approximation, the typical ~ $Q_z^{-4}$ decrease in the reflectivity curve indicates qualitatively, that the surface is smooth. In order to obtain more quantitative information from the data we compare the reflectivity profile with an optical model. In its simplest form, the model reduces to the quantum mechanical problem of a particle incident on a potential step. In our case, the optical potential has three distinct regions: vacuum, near surface, and bulk. The transition between adjacent steps calculated as an interface roughness. The reflectivity is calculated from the optical potential using a recursive definition of the reflectivity, taking into account multiple reflections [16]. The numerical description of the interface roughness follows that of Nevot and Croce [17]. From the experimental data we have calculated the optical potential shown on fig.2. It is clearly evident that, near to the surface, there is region with higher scattering power in the 1.54K data compared with those at 2.30K.

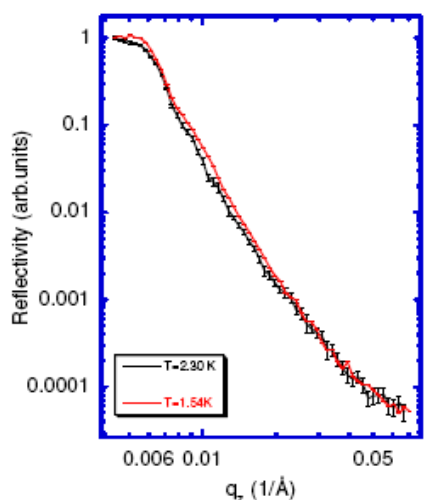

Fig.1. The reflectivity of the liquid $^4$He surface as a function of $Q$z measured at 2.3 K and 1.54K.

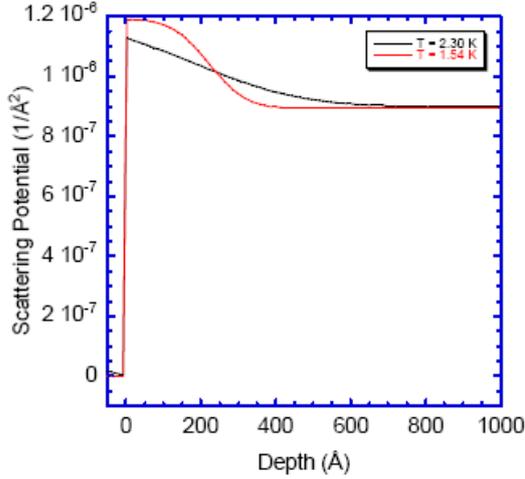

Fig.2. The scattering potential profiles obtained by numerical simulation, using the reflectivity data for 2.3 K and 1.54 K; the abscissa corresponds to vacuum below 0 Å, to the near-surface region between 0 Å and 800 Å, and to bulk liquid $^4$He above 800Å.

At 1.54K the transition from bulk to vacuum is compressed into ~400 Å whereas, at 2.30K, the transition occurs gradually over a large 800 Å distance. This behaviour may be attributable to the way in which Bose-Einstein condensation occurs in the liquid as it cools through the superfluid transition: an enhancement of Bose-Einstein condensate fraction close to the surface of the superfluid liquid [15, 18, 19] could account the alteration in the neutron scattering structure factor. Note that the optical potential is proportional to the density and scattering power of the material.

### 4. Neutron reflection from a $^3$He layer adsorbed on the liquid $^4$He surface

The measured reflectivity of a $^3$He-$^4$He liquid surface is shown in Fig. 3 at a relatively high temperature (1.75 K) where a surface layer of $^3$He would not be expected, and at a lower temperature (0.37 K). We suggest that the difference in reflectivity at low $Q_z$ is attributable to Andreev states of the $^3$He on the surface at low temperatures.

More than four decades ago A. F. Andreev predicted theoretically the existence of quantum states of $^3$He atoms on the free surface of liquid $^4$He [20]. In other words, at low temperatures, even for extremely dilute solutions, a substantial number of $^3$He atoms are adsorbed at the surface, forming a 2D Fermi liquid. The large difference in the neutron scattering cross section between the $^3$He and $^4$He isotopes, in combination with neutron reflection, is the only technique that may be able to access the few hundred Å length scales required and observe this layer of $^3$He.

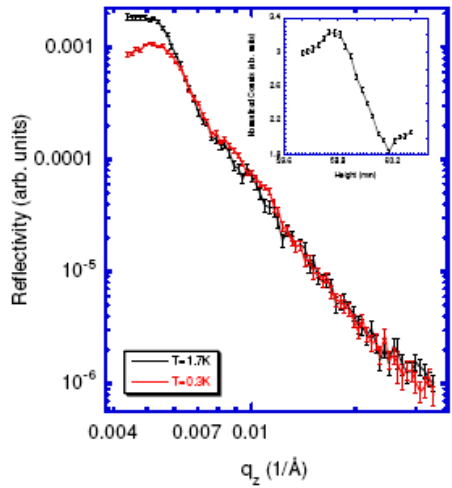

Fig.3. Reflectivity of a $^3$He (0.12%) $^4$He mixture measured at T=0.37 K (red) and T= 1.7K(black). The inset shows a height scan of the cryostat with the instrument in transmission geometry. The dip at 60.2 mm is the location of the $^3$He surface.

The different reflectivities with and without 2D Fermi layer can thus be accounted for. Experiments with different concentrations of $^3$He, and at different temperature are being planned.

## 5. Conclusions

Our neutron reflection measurements from the surface of liquid $^4$He and a very dilute $^3$He-$^4$He mixture show that the surface roughness differs above and below the lambda transition, and that the surface is smoother in the superfluid state. We observe the formation of a ~200 Å surface layer with a subtly different neutron scattering cross-section that may be attributable to an enhancement of the Bose-Einstein condensate fraction close to the helium surface. The addition of 0.12% of $^3$He brings about changes in reflectivity at low temperatures around 0.3K that would appear to be attributable to the occupation of Andreev levels.

**References**
[1] Osborne D V 1989 *J. Phys.: Condens. Matter* **1** 289
[2] Lurio L B, Rabedeau T A, Pershan P S, Isaac, Silvera F, Deutsch M, Kosowsky S D and Ocko B M 1992 *Phys. Rev. Lett*. **68** 2628
[3] Zinoveva K N 1955 *ZhETF* **29** 899
[4] Matsumoto K, Okuda Y, Suzuki M and Misawa S 2001 *J. of Low Temp. Phys.* **125** 59
[5] Abdurakhimov L V, Brazhnikov M Y, Levchenko A A, Mezhov-Deglin L P 2008 *J. Low Temp. Phys.* **150** 426
[6] Blaauwgeers R, Elstov V B, Eska G, Fine A P, Haley R P, Krusiuse M, Ruohio J J, Skrbek L and Volovik G E 2002 *Phys. Rev. Lett.* **89** 155301
[7] Volovik G E 2006 *J. Low Temp. Phys.* **145** 337
[8] Monarkha Yu and Kono K 2004 *Two-Dimensional Coulomb Liquids and Solids* Springer
[9] Atkins K R and Narahara Y 1965 *Phys. Rev.* **138** A437
[10] Grimes C C and Adams G 1976 *Phys. Rev. Lett*. **36** 145
[11] Buntar' V A, Grigoriev V N, Kirichek O I, Kovdrya Yu Z, Monarkha Yu P and Sokolov S S 1990 *J. Low Temp. Phys.* **79** 323

[12] Shirahama K, Ito S, Suto H and Kono K 1995 *J. Low Temp. Phys.* **101** 439
[13] Shirahama K, Kirichek O I and Kono K 1997 *Phys. Rev. Lett.* **79** 4218
[14] Kirichek O, Saitoh M, Kono K and Williams F I B 2001 *Phys. Rev. Letts* **86** 4064
[15] Charlton T R, Dalgliesh R M, Kirichek O, Langridge S, Ganshin A and McClintock P V E  2008 *Low Temp. Phys.* **34** 316
[16] Parratt L G 1954 *Phys. Rev.* **95(2)** 359
[17] Nevot L and Croce P 1980 *Rev. Phys. Appl.* **15** 761
[18] Griffin A and Stringari S 1996 *Phys. Rev. Lett.* **76** 259
[19] Wyatt A F G 1998 *Nature* **391** 56
[20] Andreev A F 1966 *Sov. Phys. JETP* **23** 939
[21] Lawes G, Golov A I, Nazaretski E, Mulders N and Parpia J M 2003 *Phys. Rev. Lett.* **90** 195301
[22] Zinoveva K N, Boldarev S T 1969 *ZhETF* **56** 1089